\documentclass[12pt]{article}
\usepackage{amsmath}
\usepackage{graphicx}
\usepackage{enumerate}
\usepackage{natbib}
\usepackage{url} 
\usepackage{bm}
\usepackage{amsfonts,amssymb}
\usepackage{microtype}
\usepackage{algorithm}

\usepackage[english]{babel}
\usepackage{latexsym}
\usepackage[final]{pdfpages}
\usepackage{multirow}
\usepackage{booktabs}
\usepackage{lscape}
\usepackage{fancyref}
\usepackage{epstopdf}
\usepackage[title]{appendix}

\newtheorem{lemma}{Lemma}

\newtheorem{theorem}{Theorem}

\usepackage{soul}
\newcommand{\blind}{0}

\begin{document}

\newcommand{\itt}{\textit}
\def\spacingset#1{\renewcommand{\baselinestretch}%
{#1}\small\normalsize} \spacingset{1}


\if0\blind
{
  \title{\bf The Kendall Interaction Filter for Variable Interaction Screening in Ultra High Dimensional Classification Problems}
  \author{Youssef Anzarmou\thanks{The author would like to express his deep gratitude to Professor Karim Oualkacha (Universit\'e du Qu\'ebec \`a Montr\'eal, Montreal, QC) for the support and assistance in an internship that resulted in this work.}\\
    Department of Mathematics, University of Cadi Ayyad\\ Marrakech, Morocco\\ \\
    Abdallah Mkhadri \\
    Department of Mathematics, University of Cadi Ayyad\\ Marrakech, Morocco\\
	and \\
	Karim Oualkacha \\
	Department of Mathematics, Universit\'e du Qu\'ebec \`a Montr\'eal\\ Montreal, Qu\'ebec, Canada}
  \maketitle
} \fi

\if1\blind
{
  \bigskip
  \bigskip
  \bigskip
  \begin{center}
    {\LARGE\bf The Kendall Interaction Filter for Variable Interaction Screening in Ultra High Dimensional Classification Problems}
\end{center}
  \medskip
} \fi

\bigskip
\begin{abstract}
Accounting for important interaction effects can improve prediction of many statistical learning models. Identification of relevant interactions, however, is a challenging issue owing to their ultrahigh-dimensional nature. Interaction screening strategies can alleviate such issues. However, due to heavier tail distribution and complex dependence structure of interaction effects, innovative robust and/or model-free methods for screening interactions are required to better scale analysis of complex and high-throughput data. In this work, we develop a new model-free interaction screening method, termed Kendall Interaction Filter (KIF), for the classification in high-dimensional settings. The KIF method suggests a weighted-sum measure, which compares the overall to the within-cluster Kendall's $\tau$ of pairs of predictors, to select interactive couples of features. The proposed KIF measure captures relevant interactions for the clusters response-variable, handles continuous, categorical or a mixture of continuous-categorical features, and is invariant under monotonic transformations. We show that the KIF measure enjoys the sure screening property in the high-dimensional setting under mild conditions, without imposing sub-exponential moment assumptions on the features' distributions. We illustrate the favorable behavior of the proposed methodology compared to the methods in the same category using simulation studies, and we conduct real data analyses to demonstrate its utility.
\end{abstract}

\noindent%
{\it Keywords:} Sure independence screening; Dimension reduction; Classification; Features ranking.
\vfill

\newpage
\spacingset{1.5} 
\section{Introduction}
\label{sec:intro}

High dimensional data are becoming collected in almost every field, particularly thanks to the significant improvements in information technology. The latter made it possible to inspect new phenomenons and gather data with an unprecedented accuracy leading to an abundant amount of information to be analyzed. But, with new discoveries come new challenges, most notably here is the high dimensionality. Classical data analysis methods are ill equipped in this realm either because theoretical results are not valid anymore and/or because the computational complexity simply exceeds the hardware limits. Screening methods made themselves as a reliable solution to the latter problems as they intend to select a subset of variables that contains the ``true'' relevant features with probability tending to $1$, the sure screening property in short. Thus, achieving an efficient variable selection and providing a starting point for the classical methods to operate.
\subsection{Related work}
One of the pioneering works for feature screening is the sure independence screening (SIS) approach of \cite{Fan2008}. The SIS method captures relevant main effects based on ranking the marginal correlation of each predictor with the response. SIS enjoys the sure screening property under the linear regression framework. Many endeavors have been undertaken for feature screening henceforth to improve and/or to extend the SIS approach of \cite{Fan2008}, and the methods are broadly divided into two categories: Model-based and Model-free approaches. In the model-based framework, marginal measures for feature importance rely on imposing specific model structure of the response-features relationship. To list a few \cite{Fann2010}, \cite{Fann2011} and \cite{Wang2012}, among others. Their performance is based upon the belief that the imposed working model is close to the true model. In ultrahigh-dimensional settings, however, a large number of candidate predictors are available and little information is known about the true model forms. To remedy such issues, model-free and/or robust screening procedures have been developed, including-- but not limited to-- robust rank correlation screening (RRCS) by \cite{Li2012}, distance correlation-SIS (DC-SIS) by \cite{Lidc2012}, Kolmogorov filter (KF) by \cite{Mai2013}, Pearson chi-square-SIS (PC-SIS) by \cite{Huang2014}, and mean variance-SIS (MV-SIS) by \cite{Cui2015}. Among all the cited methods, RRCS and DC-SIS handle continuous response variables while KF, PC-SIS and MV-SIS tackle classification (i.e. discrete response). 

In both categories, most SIS-type methods assume that relationship can be inferred from individual predictor-response associations. This is a key assumption of their success. To alleviate drawbacks of independence learning, multivariate (\cite{Ke2014, Kang2017}) and group (\cite{Qiu2020}) feature screening approaches have been proposed. A selective overview of feature screening in ultrahigh dimensional settings can be found in \cite{Fan2018}; a recent overview with specific focus on practical performance of feature screening procedures on the analysis of neuroimaging data is given in \cite{He2019}. 

As stated earlier, most existing SIS-type feature screening methods process predictors to capture marginal main effects, ignoring features' interplay impact on the outcome. Accounting for important interaction effects can substantially contribute in explaining the outcome total variation and might help improve the prediction of many statistical learning models. In genetics field, for instance, the study of gene-gene (i.e. epistasis) and gene-environment interactions has been a focus of research for several years (\cite{Phillips2008}, \cite{Cordell2009}); such genetic interactions play a crucial role for the etiology, prognosis and response to treatment of many complex human diseases beyond the main effects (\cite{Moore2003}). Yet, with the emerging of Multi-omics data collection (genomics, epigenomics, trancriptomics, metabolomics), the interplay between DNA methylation (epigenomics marks) and near-by SNPs (genomics markers) in influencing the patterns of gene expression (transcriptomics profiles) is a focus of many recent pharmacogenomics applications to contribute to ``precision medicine'' and treatment plans tailored to the genetic makeup of patients.

Identification of relevant interactions, however, is a challenging issue due to their ultrahigh-dimensional nature. Although feature screening interaction strategies can be a key solution to alleviate such issues, screening for interaction effects has received less consideration in the literature, but is gaining attention fast. The methods fall under the model-based/free frameworks, however, we prefer to categorize them as heredity based-/free-assumption approaches since several methods heavily rely on such an assumption for selecting interactive couples of features. 

Among the methods belonging to heredity based-assumption category, \cite{Hall2014} proposed a recursive screening procedure based on a two-steps marginal SIS-type procedure, for additive models. In linear regression framework, \cite{Hao2014} suggested an interaction forward selection based method (iFORM) where in each step a main effect or a two-way interaction is selected based on BIC criteria. Recently, \cite{Li2019} developed a stepwise conditional likelihood variable-selection approach for discriminant analysis (SODA) based on both a forward selection and a backward elimination using the extended BIC criteria. Most methods in the first category are sensitive to heredity assumption and might fail in capturing interactive features that have no main effects. 

Consequently, heredity free-assumption methods have been proposed to tackle this problem, especially for continuous outcomes. For instance, interaction pursuit (IP-SIS) proposed by \cite{Fan2016} captures interaction variables based on Pearson correlation between square transformed predictor-response variables; the idea of using such a marginal utility function is that, under linear regression framework, the correlation between the transformed predictor-response is always non zero as long as the feature is an active interaction variable. \cite{Kong2017} relied on distance correlation measure between two vectors, proposed by \cite{Szekely2007}, and extended the interaction pursuit method to multiple-response regression (IPDC). \cite{Pan2019} used predictor-response squared transformation trick within a ball-correlation (BCor) distance to screen for interactive variables, in a similar way as the IPDC approach. However, since BCor measure is a rank-based distance, it is robust to heavy-tailed distributions and does not require finite moments, compared to IPDC which relies on a Pearson-correlation distance.

In the classification framework, heredity free-assumption and/or model-free interaction screening methods are lacking. In a specific situation, where both features and outcome are categorical variables, \cite{Huang2014} extended the Pearson Chi-Square SIS (PC-SIS) marginal measure to feature interaction importance. For binary classification, \cite{Fan2015} proposed an innovated interaction screening-sparse quadratic discriminant analysis (IIS-SQDA). The method uses a transformation trick of the original $p$-dimensional feature vector to capture interactive variables in a first step followed by a sparse QDA to select important interactions and main effects. However, transformation used in IIS-SQDA relies on the estimation of sparse precision matrices in both classes, which might be a difficult problem in high-dimensional settings. Recently, joint cumulant interaction screening (JCIS) method is proposed by \cite{Reese2018}. To screen interactive features, JCIS ranks a three-way joint cumulant measure (\cite{JamesHu1991}, \cite{Nica2006}) between two features at a time and the outcome; such a measure generalizes Pearson correlation for three variables. JCIS handles both discrete and continuous outcomes and it is a model-free procedure that enjoys the sure screening property, however, it is a Pearson-correlation based method and can be sensitive to heavy-tailed distributions and requires finite second moments of the predictors and the outcome.

\subsection{Our proposal}
In all previously mentioned methods, and existing interaction methods in the literature that have not been mentioned, to our knowledge, a unified interaction screening method for \emph{classification} that
i) is model-free, ii) has the ability to process both continuous and categorical features, iii) has the sure screening property, iv) is heredity-assumption free, and v) is robust against heavy-tailed distributions, is lacking in the literature. Consequently, we develop a model-free interaction screening method meant for multiclass classification. The method is called Kendall Interaction Filter, KIF for short, and aims to select relevant couples based on ranking their KIF scores. For a pair of predictors, the KIF score is obtained as a weighted-sum of the distance between the overall and the within-cluster Kendall's $\tau$, over all clusters. Since KIF is a rank-based measure, it is non parametric, robust against heavy-tailed distributions, and invariant against monotone transformations. The KIF method can deal with both continuous and categorical predictors, either in a separated fashion or a mixed one. It enjoys the sure screening property under mild conditions. Its formula is simple, and hence its implementation is straightforward especially that empirical Kendall's $\tau$ can be obtained from almost every statistical software. Our interaction screening algorithms are implemented in the {\it KIF} \itt{R} package hosted on GitHub (\url{https://github.com/KarimOualkacha/KIF}).

The rest of the paper is organized as follows: the KIF methodology is presented in Section 2. Section 3 contains the theoretical results. Section 4 shows the behavior of KIF in both simulated and real data. Several remarks and a discussion conclude the paper.

\section{Methodology}
\label{sec:Methodology}
Consider a set of $p$ predictors, $X_{1},\dots, X_{p}$, and a label variable $Y \in \{1,..., K\}$ with $K$ an integer. \\
Our goal is to select the relevant couples of predictors for classification based on their interactions with respect to the response variable $Y$. By relevant couple, we mean two variables whose \emph{joint} behavior (i.e. concordance-based association strength at the population level) differs from one class to another, at least for two classes. Put in another way, we consider two variables as a non active couple to classification if their joint behavior or association strength is the same within all clusters. For each pair among all the possible $p(p-1)/2$ pairs, our method captures a relevant couple based on comparison of its within-class associations to the global association (independently of $Y$). We adopt Kendall's $\tau$ rank correlation measure to evaluate the association both ways: globally and with respect to each class.
Let $\Tilde{X}_1,\dots, \Tilde{X}_p$, and $\Tilde{Y}$ be independent copies of $X_1,\dots, X_p$ and $Y$, respectively. Then, for $1\leq j,l \leq p $, Kendall's $\tau$ can be expressed as follows
\begin{align}\label{tau}
	\tau(X_j,X_l) &=  \mathbb{P}\big(( X_j -  \Tilde{X}_j)( X_l -  \Tilde{X}_l)>0\big) - \mathbb{P}\big(( X_j -  \Tilde{X}_j)( X_l -  \Tilde{X}_l)<0\big)   \nonumber  \\
	&= 2 \mathbb{P}\big(( X_j -  \Tilde{X}_j)( X_l -  \Tilde{X}_l)>0\big) -1.
\end{align}
In a similar way, the rank-based association with respect to each class $k$ among the $K$ classes can be evaluated using conditional Kendall's $\tau$ as follows
\begin{equation}\label{cond/tau}
	\tau_k(X_j,X_l) = 2 \mathbb{P}\big(( X_j -  \Tilde{X}_j)( X_l -  \Tilde{X}_l)>0\ | Y=k , \Tilde{Y}=k \big) - 1.
\end{equation}
Kendall's $\tau$ is intuitive, well known by the community, best known for its robustness to monotonic transformations, implemented in most statistical software, and has a simple formula. In contrast to Pearson correlation measure, Kendall's $\tau$ is always defined, even if some variables have no finite second moment.\\
Our method then relies on the quantities introduced in (\ref{tau}) and (\ref{cond/tau}) to define the population version of the Kendall Interaction Filter score (KIF) of each couple of variables $(X_j,X_l)$ as follows
\begin{align*}
w_{j,l} = \sum_{k = 1}^{K} \pi_k |\tau_{k}(X_j,X_l) - \tau(X_j,X_l)|,    
\end{align*}
where $\pi_k = \mathbb{P}(Y = k)$ is the prior probability. The KIF measure is a rank-based non negative statistic, and thus it is less sensitive to strictly monotone transformations. KIF varies between $0$ and $1$, and has the advantage of being equal to $0$ if and only if $(X_j,X_l)$ rank-based association is independent of $Y$. It reaches its upper bound 1, for instance, in a balanced binary classification (i.e. $K=2$, $\pi_1=\pi_2=0.5$), in the extreme case where $X_j$ and $X_l$ are totally concordant in class $1$ and discordant in class $2$. In fact, in this case one has $\tau_1(X_j,X_l) = -1$ and $\tau_2(X_j,X_l) = 1$, which leads to global/marginal Kendall's $\tau(X_j,X_l) = 0$; thus, one can easily verify that $w_{j,l}=1$. Consequently, the KIF measure has the ability to capture different interactivity levels of couples of features that might impact the classification process. \\
Notice that for an interactive couple, say $(X_j,X_l)$, the class-association measures $\tau_k$'s need to behave differently between classes (at least for two classes). Thus, the KIF measure can also be interpreted as a statistic quantifying the magnitude (in absolute value) of the between-class association differences with respect to the marginal association measure. The difference in absolute value is meant to differentiate a relevant couple to classification from a couple that is highly dependent but does not contribute to the outcome $Y$; the weights $\pi_k, k=1,\ldots, K,$ adjust for the contribution of each class with respect to its size. \\

The empirical version of $w_{j,l}$ for $j,l \in \{1,...,p\}$ is defined as follows
\begin{equation}\label{kifHat}
    \hat{w}_{j,l} = \sum_{k = 1}^{K} \hat{\pi}_k |\hat{\tau}_{k}(X_j,X_l) - \hat{\tau}(X_j,X_l)|,    
\end{equation}
where $\hat{\pi}_{k} = \dfrac{1}{n}\sum_{i=1}^{n}1\!\!1\{Y_i=k\}$, and
\begin{align*}
    \hat{\tau}(X_j,X_l) &= \dfrac{4}{n(n-1)}\sum_{i<t = 1}^{n}1\!\!1\{(X_{ij} - X_{tj})(X_{il} - X_{tl})>0\} - 1, \\
	\hat{\tau}_{k}(X_j,X_l) &= \dfrac{4}{n_k(n_k-1)}\sum_{i<t = 1}^{n}1\!\!1\{(X_{ij} - X_{tj})(X_{il} - X_{tl})>0, Y_i = k, Y_t = k\} - 1,
\end{align*}
with $n_k$ the number of observations belonging to class $k$. 

To better elucidate the KIF measure functioning, next is a toy example illustrating the behavior of the empirical version of the KIF scores. The example shows that relevant pairs in the data tend to receive higher KIF scores, most likely, compared to the irrelevant couples. We generate a dataset consisting of $n=200$ observations and $p=1000$ variables. The important couples are $(X_1 , X_2)$ and $(X_3 , X_4)$, and the binary response variable $Y$ verifies $\mathbb{P}(Y=1) = \mathbb{P}(Y=0) = 0.5$. $X_1$, $X_2$, $X_3$ and $X_4$ follow a uniform distribution $\mathcal{U}[-1,1]$. Their dependence structure is expressed in Figure $1$. If for example $X_1$ is in the range $[-1, -1/3[$ and $X_2$ is in the range $[-1, -1/3[$, then $Y=1$ as shown on the left table of Figure $1$; if $X_3$ is in the range $[1/4, 1]$ and $X_4$ is in the range $[-1/4, 1/4[$, then $Y=1$ as shown in the right table of Figure $1$. The rest of the features $ X_j$, for $5\leq j \leq p$, are generated independently from a uniform distribution $\mathcal{U}[-1,1]$.\\
We generate 100 replications of the data and, in each time, we compute KIF scores for all possible $p(p-1)/2$ couples, then we compare the scores of couples $(X_1,X_2)$ and $(X_3,X_4)$ to the maximum score of all remaining pairs. The results are shown in Figure \ref{figure2}. As one can see, KIF assigns the highest scores to the important couples $(X_1,X_2)$ and $(X_3,X_4)$, in almost each time. This confirms KIF's ability to detect relevant pairs in the data.
\begin{center}
    [Figure 1 near here]
\end{center}
\begin{center}
    [Figure 2 near here]
\end{center}

Next we outline our interaction screening procedure based on the KIF measure, as follows: once all the scores $\hat{w}_{j,l}$ are obtained, we sort them decreasingly and then we choose the significant couples based on the ranking of their related scores using a prefixed threshold. More formally, we define
\begin{itemize}
	\item $\bm S := \{(j,l),\ F( Y | \bm X)\ functionally\ depends\ on\ the\ couple\ (X_j, X_l) \ rank\ association \}$ as the set of the relevant couples;
	\item $|\bm S| = s$, with $s$ much smaller than $p$ (the sparsity assumption); 
	\item $\hat{\bm S} := \{(j,l),\ \hat{w}_{j,l}>cn^{-r} \}$, where $c$ and $r$ are positive constants.
\end{itemize}
Then, our aim is to obtain an estimated set $\hat{\bm S}$ such that $\bm S \subseteq\hat{\bm S}$, with probability approaching 1. This is the focus of Section 3.\\
The threshold $cn^{-r}$ controls the number of selected pairs by the KIF method, and it can be tuned using cross-validation, in practice. However, we do note pursue this route and adopt instead the threshold $\lceil n/log(n)\rceil$ as the number of couples to choose, similar to \cite{Fan2008} and \cite{Mai2013}.

Finally, it is worthwhile to mention that there is a variety of ways to discuss and to measure dependence between the active couples and the classification outcome. Our KIF measure (and its corresponding active set $\bm S$) is, in the words of Hoeffding (\cite{hoffding1940},\cite{hoeffding1941}), a ``scale-invariant" dependence measure. Thus, although $w_{j,l} = 0$ does not necessarily imply that the couple $(X_j, X_k)$ and $Y$ are completely independent, it indicates that the rank association of $(X_j, X_k)$ is independent of $Y$. Consequently, the KIF measure captures an important form of scale-invariant concordance dependence between $(X_j, X_k)$ and $Y$, and thus it has great power of detecting complex associations under less restrictive assumptions. This is well demonstrated through exhaustive simulation studies of Section \ref{sec:Empirical study}.

\section{Theoretical results}
\label{sec:Theoretical results}
In this section, we establish the theoretical properties for the KIF method. We show that under mild conditions, our method enjoys both the sure screening property and the ranking consistency property. To this end, consider the following Lemmas, which are essential for theoretical developments.
\begin{lemma}.
	Consider two random variables $X_j$ and $X_l$, for $j,l \in \{1,...,p\}$, with $n$ observations $(X_{1,j},...,X_{n,j})$ and $(X_{1,l},...,X_{n,l})$, respectively. The Kendall's $\tau$ and its empirical version verify the following
	\begin{itemize}
		\item [i.] $\mathbb{E}\big(\hat{\tau}(X_j,X_l)\big) = \tau(X_j,X_l)$;
		\item[ii.] $\mathbb{P}\big(|\hat{\tau}(X_j,X_l)- \tau(X_j,X_l)| > \epsilon \big) \leq 2\exp\big(-n\epsilon^{2}/8\big), \quad \textrm{for all} \quad \epsilon > 0$.
	\end{itemize}
\end{lemma}

\begin{lemma}
	Consider two random variables $X_j$ and $X_l$, for $j,l \in \{1,...,p\}$, with $n$ observations $(X_{1,j},...,X_{n,j})$ and $(X_{1,l},...,X_{n,l})$, respectively. For $k=1,...,K$, the conditional Kendall's $\tau$ and its empirical version verify the following 
	\begin{itemize}
		\item [i.] $\mathbb{E}\big(\hat{\pi}_{k}\hat{\tau}_k(X_j,X_l)\ |\ Y\big) = \hat{\pi}_{k}\tau_k(X_j,X_l)$;
		\item[ii.] $\mathbb{P}\big(\hat{\pi}_{k}|\hat{\tau}_k(X_j,X_l)- \tau_k(X_j,X_l)| > \epsilon\ |\ Y \big) \leq 2\exp\big(-n\epsilon^{2}/8\big), \quad \textrm{for all} \quad \epsilon > 0$.
	\end{itemize}
\end{lemma}
Proofs of Lemmas 1 and 2 are presented in Appendix {\bf \ref{appA}} and {\bf \ref{appB}}, respectively. \\ Basically, these two Lemmas show that our estimators for both Kendall's $\tau$ and conditional Kendall's $\tau$ are consistent. Notice that, to obtain these results, we did not have to assume neither sub-exponential bounds nor normality assumptions on the covariates' distributions. This makes our method more flexible.\\ 
Next are the conditions that we assume:
\begin{itemize}
	\item [(C1)] There exist two positive constants $c_1$ and $c_2$ such that \\$c_1/K \leq \min\limits_{1 \leq k \leq K } \pi_k \leq \max\limits_{1 \leq k \leq K } \pi_k \leq c_2/K$. 
	\item [(C2)] There exist positive constants $c > 0$ and $0 \leq r < 1/2$ such that $\min\limits_{j,l \in \bm S}w_{j,l} > 2cn^{-r} $.
	\item [(C3)] Assume the number of classes $K = O(n^d)$ for $0 \leq d <\dfrac{1}{2} - r$.
	\item [(C4)] Assume $\log(p) = o(n^{1-2(r+d)})$. 
\end{itemize}
Conditions (C1)--(C4) are standard assumptions for most of the screening methods; they are similar to the ones adopted in \cite{Cui2015} and \cite{Pan2019}. Condition (C1) assumes that  the proportion of each class is neither too small nor too large. Condition (C2) implies that KIF minimal true signal is positive. Condition (C3) allows the number of groups to grow as $n$ increases. In condition (C4), we let $p$ to be at an exponential order of $n$.

Next, we present our first main result; the sure screening property.  
\begin{theorem}(Sure Screening property).\\
	Suppose conditions (C1)--(C4) are verified. There exists a positive constant b depending on c, $c_1$, and $c_2$, such that,
	\begin{itemize}
		\item [i.] $
		\mathbb{P}\Big(\max_{1\leq j,l\leq p}|\hat{w}_{j,l} - w_{j,l}|\ \leq cn^{-r}  \Big) \geq 1 - O\bigg(p^{2
		}n^{d}\exp\Big( -c^{2}n^{1-2(r+d)}/72b^{2}\Big)\bigg).$
		\item [ii.]$
		\mathbb{P}\Big(\bm S \subseteq\hat{\bm S}  \Big) \geq 1 - O\bigg(sn^{d}\exp\Big( -c^{2}n^{1-2(r+d)}/72b^{2}\Big)\bigg).$
	\end{itemize}
	\begin{flushright}
		$\square$
	\end{flushright}
\end{theorem}
Proof of Theorem 1 is postponed to Appendix {\bf \ref{appC}}. \\
Theorem $1$ shows that KIF is consistent in estimating the set of true significant couples. It assures that, with probability tending to $1$, the estimated set of features contains the true significant couples. In Theorem $1$ we only proved that $\bm S \subseteq\hat{\bm S}$; however, the reverse inclusion, $\hat{\bm S} \subseteq \bm S$, can be deduced from using an important property of the KIF method which states that the KIF measure is equal to $0$ if and only if $(X_j,X_l)$ rank-based association and $Y$ are independent, for $j,l \in \{1, ..., p\}$.\\
Next, we establish the ranking consistency property. First, consider the following condition
\begin{itemize}
	\item [(C5)] $\liminf\limits_{p\rightarrow\infty} \{\min\limits_{j,l\in \bm S}w_{j,l} - \max\limits_{j,l\in \bm S^{c}}w_{j,l} \} \geq c_3$, where $c_3 > 0$ is a constant.
\end{itemize}
Condition (C5) requires the relevant and irrelevant couples to be distinguishable at the population level. 
\begin{theorem}Ranking Consistency property. \\
	Assume conditions (C1)--(C5) to hold for $K^2\log(n)/n = o(1)$. Then, $\liminf\limits_{n\rightarrow\infty} \{\min\limits_{j,l\in \bm S}\hat{w}_{j,l} - \max\limits_{j,l\in \bm S^{c}}\hat{w}_{j,l} \} > 0$ a.s.
	\begin{flushright}
		$\square$
	\end{flushright}
\end{theorem}
Proof of Theorem 2 is outlined in Appendix {\bf \ref{appD}}.\\
The ranking consistency property is a strong property compared to the sure screening property. It states that the relevant couples always have their KIF scores bigger than the irrelevant pairs. Consequently, in an ideal situation where the correct number of the true couples is known, we will be able to recover $\bm S$ perfectly.
\section{Empirical study}
\label{sec:Empirical study}
\subsection{Simulation study}
In what follows, we conduct a detailed empirical study based on five simulation settings to show the good behavior of KIF with respect to five competing methods: JCIS, BCOR-SIS, IP-SIS, SODA and PC-SIS. In all settings we set the sample size and the number of features to be $n = 200$ and $p=500$, respectively. In each simulation scenario, we generate 100 random data sets, and we assess the methods' performance based on the selection rate statistic, defined as the number of times the methods select the significant couples among the $\lceil n/log(n)\rceil$ first selected couples, in $100$ replications. The simulated scenarios model the predictors dependence in various ways to allow fair judgment of KIF behavior against the other competing methods. While KIF, JCIS, SODA and PC-SIS select directly the important couples, BCOR-SIS and IP-SIS operate differently and select the predictors that are either main effects or components in significant couples. As a result, for both BCOR-SIS and IP-SIS we first select $q$ predictors and then construct all the possible couples based on the chosen predictors. For fair comparison with the other methods, the number $q$ is chosen such that $q(q-1)/2 \simeq \lceil n/log(n)\rceil$.

\subsubsection{Simulation setting 1}
 By this example, we aim to inspect the behavior of the methods with respect to heredity. We first simulate four standard interaction scenarios following a logistic model as follows 
 \begin{itemize}
	\item [$i$.] $\log\left(\dfrac{\mathbb{P}(Y=1 |\bm X)}{1 - \mathbb{P}(Y=1 |\bm X)}\right) = 2X_1 + 2X_2 + X_1X_2$;
	\item [$ii$.] $\log\left(\dfrac{\mathbb{P}(Y=1 |\bm X)}{1 - \mathbb{P}(Y=1 |\bm X)}\right) = X_1 + X_5 + X_1X_2$;
	\item [$iii$.] $\log\left(\dfrac{\mathbb{P}(Y=1 |\bm X)}{1 - \mathbb{P}(Y=1 |\bm X)}\right) = X_5 + X_{10} + X_1X_2$;
	\item [$iv$.] $\log\left(\dfrac{\mathbb{P}(Y=1 |\bm X)}{1 - \mathbb{P}(Y=1 |\bm X)}\right) = X_1X2$. 
\end{itemize}
The vector $\bm X = (X_1,...,X_p)^\top$ follows a multivariate normal distribution, $\mathcal{N}(\bm 0_p,\Sigma)$, with $\Sigma = (0.2^{|j-l|})_{1\leq j,l \leq p}$.\\
We compare KIF with JCIS, SODA, BCOR-SIS and IP-SIS. Of note, BCOR-SIS and IP-SIS are conceived to handle continuous outcomes. Thus, to allow for comparison with IP-SIS and BCOR-SIS, we use the linear mean function of the four models as a continuous variable for these two methods (i.e right-hand-side part of the 4 models), before applying the logistic transformation to generate the binary response. The four scenarios represent different levels of heredity going from high in scenario $i$ and down to completely no heredity in scenario $iv$. Results of this scenario can be found in Table $1$.
Based on Table $1$, globally, KIF outperforms all methods, followed by IP-SIS, and the two methods are consistent whether the heredity level is high, moderate or low. On the other hand, SODA performs relatively well in Scenarios $1$ and $2$, however the method shows substantial decreases of the selection rate in Scenarios $3$ and $4$. This indicates that SODA is highly dependent on the heredity assumption. JCIS and BCOR-SIS fail in Scenarios $1$ and $3$, respectively. 
\begin{center}
    [Table 1 near here]
\end{center}

\subsubsection{Simulation setting 2}
The aim here is to test the four methods robustness against monotone transformations.
We replicate the $4$ scenarios of Setting 1, however, all the methods are fitted using the transformed predictors, $W_j = \exp(X_j)$, $j=1,\ldots,p$. Similar to Setting 1, before applying the logistic transformation to generate the binary response, the linear mean function of the four models is used as a continuous variable for the IP-SIS and BCOR-SIS methods. We again compare KIF with JCIS, SODA, BCOR-SIS and IP-SIS. Results are summarized in Table $2$. KIF, BCOR-SIS and SODA scores are similar to those obtained in Setting 1 for the four scenarios, which can prove that the three methods are insensitive to monotone transformations. In contrast, a relative drop of the selection rate can be observed in IP-SIS performance compared to its results in Setting 1, which suggests that the method is relatively affected by monotone transformations. On the other hand, JCIS shows a poor performance and clearly the method is highly affected by monotone transformations.
\begin{center}
    [Table 2 near here]
\end{center}

\subsubsection{Simulation setting 3}
In this setting, we consider a two-class Gaussian classification model where the two classes have equal means, however, interactive couples are assumed to have different correlations in the two classes. We assume two interactive couples with moderate and high interaction effects on the two-class response variable. Our goal is to investigate the methods ability for detecting moderate-to-high levels of interaction effects. 

Let the features vector, ${\bm X}$, be distributed following a mixture of two multivariate normal distributions, as follows
\[ {\bm X} | Y=1 \sim N(\bm 0_p, \Sigma_1),\ \ \ \textrm{and} \ \ \ {\bm X} | Y=0 \sim N(\bm 0_p, \Sigma_2),
\]
where
\[\Sigma_{1,{(j_1,j_2)}}=
\begin{cases}
1 & \text{if $j_1=j_2$ } \\
-0.8 & \text{if $(j_1,j_2)\in \{(3,4),(4,3)\}$ }  \\
0.2 & \text{elsewhere}, 
\end{cases}\] \\
and 
\[\Sigma_{2,{(j_1,j_2)}}=
\begin{cases}
1 & \text{if $j_1=j_2$ } \\
0.8 & \text{if $(j_1,j_2)\in \{(1,2),(3,4),(4,3),(2,1)\}$ }  \\
0.2 & \text{elsewhere}.
\end{cases}\]
Thus, $(X_1,X_2)$ and $(X_3,X_4)$ have moderate and high impact on the classification outcome, respectively. \\
Of note, generating a latent continuous variable in this scenario is not straightforward. Thus, BCOR-SIS and IP-SIS are not considered hereafter. The results are given in Table $3$. Both KIF and JCIS behave accurately in selecting the relevant couples with a small advantage to the KIF method. Although couple $(X_1,X_2)$ is difficult to select compared to $(X_3,X_4)$, KIF reaches a significant high selection rate ($90\%$) for this couple. SODA, on the other hand, fails badly and never select the relevant couples among the $100$ replications. In fact, SODA heavily relies on the heredity assumption; such a condition is violated in this simulation setting since the interactive features have no main effects on the classification outcome.
\begin{center}
    [Table 3 near here]
\end{center}

\subsubsection{Simulation setting 4}
In this scenario, we test the ability of the KIF method to differentiate important couples to the response variable from pairs that are marginally highly correlated but are not relevant to the outcome.

Let the features vector, ${\bm X}$, be distributed following a mixture of multivariate normal distributions, as follows
\[
\bm X | Y=1 \sim N(\bm 0_p, \Sigma_1)\ \ \ \textrm{and}\ \ \ \bm X | Y=0 \sim N(\bm 0_p, \Sigma_2),
\]
where 
\[\Sigma_{1,{(j_1,j_2)}}=
\begin{cases}
1 & \text{if $j_1=j_2$ } \\
0.8 & \text{if $(j_1,j_2)\in \{(1,2),(2,1),(3,4),(4,3)\}$ }  \\
0.2 & \text{elsewhere},
\end{cases}\]
and 
\[\Sigma_{2,{(j_1,j_2)}}=
\begin{cases}
1 & \text{if $j_1=j_2$ } \\
0.8 & \text{if $(j_1,j_2)\in \{(3,4),(4,3)\}$ }  \\
0.2 & \text{elsewhere}.
\end{cases}\] 
Thus, $(X_1,X_2)$ is important to $Y$ and $(X_3,X_4)$ has high marginal correlation but not relevant to $Y$. Results of this scenario are detailed in Table $4$. Similar to Scenario $3$, our method does outperform the JCIS method in selecting the important pair $(X_1,X_2)$. On the other hand, both KIF and JCIS neglect the highly correlated couple, $(X_3,X_4)$, that is not relevant for $Y$. Once again, SODA fails to select the relevant couple, $(X_1,X_2)$, due to its dependence on the heredity assumption and the fact that both $X_1$ and $X_2$ do not contribute as main effects.
\begin{center}
    [Table 4 near here]
\end{center}

\subsubsection{Simulation setting 5}
Next is a simulation setting borrowed from \cite{Reese2018}, which models the interaction in a complex way and only focuses on binary features. \\
First, we generate the outcome $Y$ following a Bernoulli distribution with $\mathbb{P}(Y=1) = \pi_1$ and $\mathbb{P}(Y=0) = \pi_0$. Then, four interactive couples, $\{(X_{2j-1},X_{2j}), j=1,2,3,4\}$, are generated according to the following conditional probabilities 
\[
\mathbb{P}\big(X_{2j-1} = 1 | Y = k\big) = \theta_{kj}, \text{ for } k=0, 1,
\]
\[
\mathbb{P}\big(X_{2j}=1|Y=k, X_{2j-1}=0 \big) = 0.6 \,1\!\!1(\theta_{kj} > 0.5) + 0.4\, 1\!\!1(\theta_{kj} \leq 0.5),
\]
and 
\[
\mathbb{P}\big(X_{2j}=1|Y=k, X_{2j-1}=1 \big) = 0.95 \, 1\!\!1(\theta_{kj} > 0.5) + 0.05 \, 1\!\!1(\theta_{kj} \leq 0.5).
\]
The values of $\theta_{kj}$, $j=1,2,3,4$, are given in Table $5$.
\begin{center}
    [Table 5 near here]
\end{center}
The remaining variables $X_j$'s, for $9\leq j \leq p$, are generated as follows
\[
\mathbb{P}(X_j=1) = \mathbb{P}(X_j=0) = 0.5.
\]
We consider three scenarios:
\begin{itemize}
	\item A balanced scenario where $\pi_0 = \pi_1 = 0.5$;
	\item An unbalanced scenario where $\pi_0 = 0.7$ and $\pi_1 = 0.3$;
	\item An unbalanced scenario where $\pi_0 = 0.3$ and $\pi_1 = 0.7$.
\end{itemize}

\begin{center}
    [Table 6 near here]
\end{center}
Since all the features are categorical, PC-SIS can be applied here. Consequently, we use JCIS, SODA and PC-SIS to judge KIF performance. The results are shown in Table $6$. KIF behaves well in all scenarios. Although its selection rate decreases when the class proportions are unbalanced, KIF maintains relatively good performance in all scenarios with a selection rate that varies between $0.62$ and $1$. This indicates that KIF is less sensitive to class proportion changes. In contrast, PC-SIS is clearly affected by changes in proportions and produces unstable results with selection rate values ranging from $0.20$ to $0.80$. On the other hand, JCIS  performance is very good in the second scenario, in which the method reaches selection rate of $1$, for the four relevant couples. However, the method fails to capture the significant couples in the first and third scenarios, which suggests that JCIS is highly affected by changes of the class proportions. Lastly, SODA performs badly in all three scenarios with selection rate scores below $0.3$, regardless of the proportions changes, suggesting that SODA is not suitable for these scenarios.

\subsection{Real data analysis}
This section aims to demonstrate the KIF approach utility in real data analyses. It illustrates KIF performance on $4$ high-dimensional real datasets available through the \itt{R} package software, \emph{datamicroarray}, and described in Table $7$. 
\begin{center}
    [Table 7 near here]
\end{center}
For each dataset, we first perform a univariate screening step to reduce noise variables. More precisely, we screen the features marginally, based on their variances; i.e. the features are first ordered from highly to slightly dispersed variable, then $20\%$ less dispersed predictors are eliminated. We then run the KIF approach for all pair combinations of the remaining predictors. \\
To validate if the findings based on the KIF approach are not artifact, one can assess the prediction power of a classification method (e.g. logistic regression) with and without the KIF selected features. However, the results from such a procedure might be prediction-method dependent. Thus, we adopt the permutation-based procedure of \cite{Ng2005BivariateVS} to assess strength of KIF statistic in selecting relevant couples of features.\\
In order to mimic the distribution of the KIF statistic under the null hypothesis of no interaction effects between a couple, say $(j,l)$, and $Y$, we permute the outcome randomly while ignoring the two class structure; any permutation of the labels for the two outcome classes represents a random occurrence of the data under the null hypothesis of no association. We repeat this procedure $T$ times, and for each $t$, we calculate the permutation-based KIF score, $\hat{w}_{j,l}^{(t)}$, from the permuted data. Finally, the empirical significance level (i.e. $\emph{p-value}$) of the KIF statistic is approximated as follows
\[
p_{j,l}= \dfrac{\sum_{t=1}^{T}1\!\!1[\hat{w}_{j,l}^{(t)}\geq \hat{w}_{j,l}] }{T},
\]
where $\hat{w}_{j,l}$ is the observed KIF score. We set the number of permutations to $T = 100000$ to allow for more accurate $\emph{p-values}$.

Table $8$ describes the obtained results for the couples with first five higher KIF scores (i.e. KIF's most likely relevant couples) and their associated $\emph{p-values}$, for each dataset. Interestingly, Table $8$ shows very small $\emph{p-values}$ for all KIF selected couples, and this result is consistent for all analyzed datasets. This demonstrates the KIF-measure significance-level strength, and also indicates that indeed the KIF selected couples are most likely to be important to the classification task. One can also observe that some features are selected in more than one couple, such as the predictor ``$1129$'' for Alon and the feature ``$115$'' for Yeoh dataset. This indicates that such variables can be very active (i.e. \emph{super/hub} features) in a way that they might have high impact on the classification outcome.
\begin{center}
    [Table 8 near here]
\end{center}
Note though that a practitioner is better suited to judge the obtained \emph{p-values} depending on the nature of the data and the goal of the analysis.

\subsection{Computational complexity}
The KIF algorithm calculates the statistic $\hat{w}_{j,l}$, given in (\ref{kifHat}), for each pair $(j,l)$ of the $p(p-1)/2$ available pairs, which implies computing Kendall's $\tau$ $(K+1)$ times, for each couple. For a sample size $n$, Kendall's $\tau$ computational complexity is $O(n^{2})$. Thus, the KIF algorithm can be computationally challenging, especially when dealing with a large sample-size data. To circumvent KIF computational burden issues, we have implement the KIF algorithm with Kendall's $\tau$ calculated based on the \itt{R} package software \emph{ccaPP} (\cite{Alfons2016}), which reduces Kendall's $\tau$ computational complexity to $O(n\log n)$. Furthermore, the KIF statistic can be calculated independently for each pair. Present-day computers have multiple cores per processor (CPU), with each core proceeds as a single computational unit. Consequently, by distributing the set of the $p(p-1)/2$ couples to chunks over these CPU cores and running them in parallel, a significant speed-up can be obtained. For instance, running KIF on a dataset with $n=200$ observations and $p=1000$ predictors takes about $14$ seconds to compute all the $p(p-1)/2=499500$ scores, using a \emph{i5 9300H} processor equipped with $16$ Gb of memory and $4$ cores.

\section{Discussion}
\label{sec:Discussion} 
In this paper, we proposed a new interaction screening method  based on Kendall's $\tau$ and meant for multi-class classification. KIF is a model-free method and can handle both continuous and categorical variables either separated or mixed. It is robust against monotonic transformations. Under mild conditions, and without assuming finite sub-Gaussian moments of the predictors, we demonstrated that KIF enjoys the sure screening property. Through simulation studies and real datasets, we showed that KIF indeed has the ability to capture the relevant couples for a classification outcome. Lastly, we provided hints to optimize the method and reduce its computational burden. \\
Although we focused on Kendall's $\tau$ dependence measure in this work, the KIF measure can be seen as a general framework rather than a specific method. In other words, other rank-based, robust and/or model-free dependence measures can be calculated marginally and within classes and can replace Kendall' $\tau$ measure in the KIF statistic. For instance, the Gini index (\cite{Ceriani2012}), ball correlation (\cite{Pan2018}), or distance correlation (\cite{Szekely2007}) measures can be suitable alternatives to Kendall's $\tau$. \\
One would argue that our method is limited to only categorical response variables. However, the method can be extended to handle continuous response variables using a slicing-trick technique to categorize the continuous outcome first, then apply the KIF framework to the corresponding multi-class outcome. The slicing technique has been used by \cite{Mai2015} to extend the Kolmogorov Filter method to capture features' marginal main effects for a continuous response variable. 
However, optimal choice of categorization cut-points (i.e. slicing/partition scheme) is not a straightforward task. To circumvent such an issue, \cite{Mai2015} proposed a fusion scheme in which information is aggregated from different partitions and a fused Kolmogorov Filter statistic is calculated based on the fused partition, for each predictor. Adaptation of the slicing-trick procedure to the KIF framework may thus hold promise as a good model-free alternative of existing methods dealing with interaction screening for continuous response variables. However, the KIF fusion extension is not straightforward either for implementation or for theoretical development, and is left for future research investigation.

\appendix
\section{Proof of Lemma 1}\label{appA}

\setcounter{equation}{0}
\numberwithin{equation}{section}
\makeatletter 
\newcommand{\section@cntformat}{Appendix \thesection:\ }
\makeatother

i. Trivial.\\
ii. For $\bm v_1,...,\bm v_n \in \mathbb{R}^2$, consider the function
\[
g(\bm v_1,...,\bm v_n) = \dfrac{4}{n(n-1)} \sum_{i<t=1}^{n}1\!\!1\big\{(v_{i1} - v_{t1})(v_{i2} - v_{t2})>0\big\}.
\]
Suppose we change only one component, say for example $\bm v_{m}$ in the place of $\bm v_{n}$ for simplicity. We have
\begin{align*}
	&\bigg|g(\bm v_1,...,\bm v_n) - g(\bm v_1,...,\bm v_{m})      \bigg| \\
	= &\bigg| \dfrac{4}{n(n-1)}\Big[\sum_{i<t=1}^{n-1}1\!\!1\{(v_{i1} - v_{t1})(v_{i2} - v_{t2})>0\} + \sum_{i=1}^{n-1}1\!\!1\{(v_{i1} - v_{n1})(v_{i2} - v_{n2})>0\}  \Big] \\
	& - \dfrac{4}{n(n-1)}\Big[\sum_{i<t=1}^{n-1}1\!\!1\{(v_{i1} - v_{t1})(v_{i2} - v_{t2})>0\} + \sum_{i=1}^{n-1}1\!\!1\{(v_{i1} - v_{m1})(v_{i2} - v_{m2})>0\}  \Big]  \bigg| \\
	= &\bigg| \dfrac{4}{n(n-1)} \sum_{i=1}^{n-1} \Big[1\!\!1\{(v_{i1} - v_{n1})(v_{i2} - v_{n2})>0\} - 1\!\!1\{(v_{i1} - v_{m1})(v_{i2} - v_{m2})>0\}   \Big] \bigg| \\
	\leq & 4/n .
\end{align*}
Consequently, the function $g$ verifies the bounded differences inequality (\cite{McDiarmid2013}). As a result, for $\{\bm V_i,i=1,...,n\}$ an independent and identically distributed random sample of size $n$, where $\bm V_i = (V_{i1},V_{i2})$, and applying McDiarmid inequality, one has 
\[
\mathbb{P} \big(|g(\bm V_1,...,\bm V_n)- \mathbb{E}g(\bm V_1,...,\bm V_n)| > \epsilon \big) \leq 2 \exp\big(-n\epsilon^{2}/8\big).
\]
Set $X = (V_{11},...,V_{n1})^{t}$ and $X^{'} = (V_{12},...,V_{n2})^{t}$. Then, we have $g(\bm V_1,...,\bm V_n) = \hat{\tau}(X,X^{'})$. Hence, the desired result.
\begin{flushright}
	$\square$
\end{flushright}

\section{Proof of Lemma 2}\label{appB} 
i. Trivial.\\
ii. For $\bm v_1,...,\bm v_n \in \mathbb{R}^3$ and $k\in \{1,...,K\}$, consider the functions
\[
g_k(\bm v_1,...,\bm v_n) = \dfrac{4}{n(n_k-1)} \sum_{i<t=1}^{n}1\!\!1\{(v_{i1} - v_{t1})(v_{i2} - v_{t2})>0, v_{i3} = k, v_{t3} = k\},
\]
with $n_k$ the number of observations that fall into class $k$. Suppose we change only one component, say for example $\bm v_m$ in the place of $\bm v_n$ for simplicity. We have
\begin{align*}
	&\bigg|g_k(\bm v_1,...,\bm v_n) - g_k(\bm v_1,...,\bm v_m)      \bigg| \\
	= &\bigg| \dfrac{4}{n(n_k-1)}\Big[\sum_{i<t=1}^{n-1}1\!\!1\{(v_{i1} - v_{t1})(v_{i2} - v_{t2})>0, v_{i3} = k, v_{t3} = k\}\\
	& + \sum_{i=1}^{n-1}1\!\!1\{(v_{i1} - v_{n1})(v_{i2} - v_{n2})>0, v_{i3} = k, v_{t3} = k\}  \Big] \\
	& - \dfrac{4}{n(n_k-1)}\Big[\sum_{i<t=1}^{n-1}1\!\!1\{(v_{i1} - v_{t1})(v_{i2} - v_{t2})>0, v_{i3} = k, v_{t3} = k\} \\
	& + \sum_{i=1}^{n-1}1\!\!1\{(v_{i1} - v_{m1})(v_{i2} - v_{m2})>0, v_{i3} = k, v_{t3} = k\}  \Big]  \bigg| \\
	= &\bigg| \dfrac{4}{n(n_k-1)} \sum_{i=1}^{n-1} \Big[1\!\!1\{(v_{i1} - v_{n1})(v_{i2} - v_{n2})>0, v_{i3} = k, v_{t3} = k \} \\
	&- 1\!\!1\{(v_{i1} - v_{m1})(v_{i2} - v_{m2})>0, v_{i3} = k, v_{t3} = k\}   \Big] \bigg| \\
	\leq &4/n .
\end{align*}
Consequently, the functions $g_k$ verify the bounded differences inequality (\cite{McDiarmid2013}). As a result, for $\{\bm V_i,i=1,...,n\}$ an independent and identically distributed random sample of size $n$, with $\bm V_i = (V_{i1},V_{i2},V_{i3})$ and set $Y = (V_{13},...,V_{n3})^{t}$, applying McDiarmid inequality, one has
\[
\mathbb{P} \big(\big|g_k(\bm V_1,...,\bm V_n)- \mathbb{E}(g_k(\bm V_1,...,\bm V_n)\ |\ Y)\big| > \epsilon\ |\ Y\big) \leq 2 \exp\big(-n\epsilon^{2}/8\big).
\]
Set $X = (V_{11},...,V_{n1})^{t}$ and $X^{'} = (V_{12},...,V_{n2})^{t}$. Then, we have $g_k(\bm V_1,...,\bm V_n) = \hat{\pi}_{k}\hat{\tau}_k(X,X^{'})$. Hence, the desired result.
\begin{flushright}
	$\square$
\end{flushright}

\section{Proof of Theorem 1}\label{appC}
First we prove $i$.\\
For each $(j,l) \in \{1,\dots,p\}^2$, we have
\begin{align*}
	&\big|\hat{w}_{j,l} - w_{j,l}\big| \\
	= &\Big|\sum_{k = 1}^{K} \hat{\pi}_k \big|\hat{\tau}_{k}(X_j,X_l)-\hat{\tau}(X_j,X_l)\big| - \sum_{k = 1}^{K} \pi_k \big|\tau_{k}(X_j,X_l)-\tau(X_j,X_l)\big|\Big|  \\
	= &\Big|\sum_{k = 1}^{K} \hat{\pi}_k \big|\hat{\tau}_{k}(X_j,X_l)-\hat{\tau}(X_j,X_l)\big| -\sum_{k = 1}^{K} \hat{\pi}_k \big|\tau_{k}(X_j,X_l)-\tau(X_j,X_l)\big| \\
	&   + \sum_{k = 1}^{K} \hat{\pi}_k \big|\tau_{k}(X_j,X_l)-\tau(X_j,X_l)\big| -\sum_{k = 1}^{K} \pi_k \big|\tau_{k}(X_j,X_l)-\tau(X_j,X_l)\big|\Big|   \\ 
	\leq & \sum_{k = 1}^{K} \hat{\pi}_k \Big|\big|\hat{\tau}_{k}(X_j,X_l)-\hat{\tau}(X_j,X_l)\big| - \big|\tau_{k}(X_j,X_l)-\tau(X_j,X_l)\big|\Big| \\
	& +  \sum_{k = 1}^{K} \big|\tau_{k}(X_j,X_l)-\tau(X_j,X_l)\big| \big|\hat{\pi}_k - \pi_k \big| \\
	\leq & \sum_{k = 1}^{K} \big|\tau_{k}(X_j,X_l)-\tau(X_j,X_l)\big| \big|\hat{\pi}_k - \pi_k \big| + \sum_{k = 1}^{K} \hat{\pi}_k\big|\hat{\tau}(X_j,X_l)-\tau(X_j,X_l)\big|\\
	& + \sum_{k = 1}^{K} \hat{\pi}_k\big|\hat{\tau}_{k}(X_j,X_l)-\tau_{k}(X_j,X_l)\big|\\
	\leq & 2\sum_{k = 1}^{K}\big|\hat{\pi}_k - \pi_k \big| + \big|\hat{\tau}(X_j,X_l)-\tau(X_j,X_l)\big| + \sum_{k = 1}^{K} \hat{\pi}_k\big|\hat{\tau}_{k}(X_j,X_l)-\tau_{k}(X_j,X_l)\big|,
\end{align*}
where we have used $\sum_{k = 1}^{K}\hat{\pi}_k = 1$, $\tau_{k}(X_j,X_l) \leq 1$ and $\tau(X_j,X_l) \leq 1$ in the last line. Then
\begin{align}\label{eqA1+2+3}
& \mathbb{P}\Big(\big|\hat{w}_{j,l} - w_{j,l}\big|\ > \epsilon \Big) \nonumber \\
&\leq  \mathbb{P}\Big(2\sum_{k = 1}^{K}\big|\hat{\pi}_k - \pi_k \big| + \big|\hat{\tau}(X_j,X_l)-\tau(X_j,X_l)\big| + \sum_{k = 1}^{K} \hat{\pi}_k\big|\hat{\tau}_{k}(X_j,X_l)-\tau_{k}(X_j,X_l)\big| > \epsilon \Big) \nonumber \\
&\leq \mathbb{P} \Big(2\sum_{k = 1}^{K}\big|\hat{\pi}_k - \pi_k \big| > \epsilon/3\Big) + \mathbb{P} \Big(\big|\hat{\tau}(X_j,X_l)-\tau(X_j,X_l)\big| > \epsilon/3\Big) \nonumber \\
& + \mathbb{P} \Big(\sum_{k = 1}^{K} \hat{\pi}_k \big|\hat{\tau}_{k}(X_j,X_l)-\tau_{k}(X_j,X_l)\big| > \epsilon/3\Big) \nonumber \\
&\leq A_1 + A_2 + A_3\ ,
\end{align}
with 
\begin{itemize}
	\item [] $A_1 = \mathbb{P} \Big(2\sum_{k = 1}^{K}\big|\hat{\pi}_k - \pi_k \big| > \epsilon/3\Big)$.
	\item [] $A_2 = \mathbb{P} \Big(\big|\hat{\tau}(X_j,X_l)-\tau(X_j,X_l)\big| > \epsilon/3\Big)$.
	\item [] $A_3 = \mathbb{P} \Big(\sum_{k = 1}^{K} \hat{\pi}_k \big|\hat{\tau}_{k}(X_j,X_l)-\tau_{k}(X_j,X_l)\big| > \epsilon/3\Big)$.
\end{itemize}
For the quantity $A_1$, we have
\begin{align*}
	A_1 &= \mathbb{P} \Big(2\sum_{k = 1}^{K}\big|\hat{\pi}_k - \pi_k \big| > \epsilon/3\Big) \\
	&\leq \mathbb{P} \Big(\max_{k}\big|\hat{\pi}_k - \pi_k \big| > \epsilon/6K\Big) \\
	&\leq \mathbb{P} \Big(\bigcup_{k}\{\big|\hat{\pi}_k - \pi_k \big| > \epsilon/6K\}\Big)\\
	&\leq \sum_{k = 1}^{K} \mathbb{P} \Big(\big|\hat{\pi}_k - \pi_k \big| > \epsilon/6K\Big).
\end{align*}
Hence, using Hoeffding's inequality we get
\begin{align}\label{eqA1}
A_1 &\leq 2\sum_{k = 1}^{K} \exp\Big(-2n\big(\epsilon/6K\big)^{2} \Big) \nonumber \\
&\leq 2K \exp\Big(-n\epsilon^{2}/18K^{2} \Big).  
\end{align}
As for the quantity $A_2$, using $Lemma\ 1$ we get
\begin{align}\label{eqA2}
	A_2 &= \mathbb{P} \Big(\big|\hat{\tau}(X_j,X_l)-\tau(X_j,X_l)\big| > \epsilon/3\Big) \nonumber \\
	&\leq 2\exp\Big(\dfrac{-n\big(\epsilon/3\big)^{2}}{8}\Big) \nonumber \\
	&= 2\exp\Big(-n\epsilon^{2}/72\Big).
\end{align}
And finally for the quantity $A_3$, we have
\begin{align*}
	A_3 &= \mathbb{P} \Big(\sum_{k = 1}^{K} \hat{\pi}_k \big|\hat{\tau}_{k}(X_j,X_l)-\tau_{k}(X_j,X_l)\big| > \epsilon/3\Big) \\
	&\leq \sum_{k = 1}^{K}\mathbb{P} \Big( \hat{\pi}_k \big|\hat{\tau}_{k}(X_j,X_l)-\tau_{k}(X_j,X_l)\big| > \epsilon/3K\Big) \\
	&\leq  \sum_{k = 1}^{K}\mathbb{E}_Y \bigg[\mathbb{P} \Big(\big|\hat{\pi}_k\hat{\tau}_{k}(X_j,X_l)-\hat{\pi}_k\tau_{k}(X_j,X_l)\big| > \epsilon/3K\ \big|\ Y\Big)\bigg].
\end{align*}
Thus applying $Lemma\ 2$ we obtain
\begin{align}\label{eqA3}
A_3 &\leq 2\sum_{k = 1}^{K} \exp\Big( \dfrac{-n\big(\epsilon/3K\big)^{2}}{8}\Big) \nonumber \\ 
&\leq 2 K\exp\Big( -n\epsilon^{2}/72K^{2}\Big).
\end{align}
Hence, combining (\ref{eqA1}), (\ref{eqA2}) and (\ref{eqA3}) along with (\ref{eqA1+2+3}) we obtain
\begin{equation*}
\mathbb{P}\Big(\big|\hat{w}_{j,l} - w_{j,l}\big|\ > \epsilon\Big) \leq 2K \exp\Big(-n\epsilon^{2}/18K^{2} \Big) + 2\exp\Big( -n\epsilon^{2}/72\Big) + 
2K\exp\Big( -n\epsilon^{2}/72K^{2}\Big).
\end{equation*}
As a result, we get
\begin{equation}\label{c.5}
\begin{split}
\mathbb{P}\Big(\max_{1\leq j,l\leq p}\big|\hat{w}_{j,l} - w_{j,l}\big|\ > \epsilon\Big) \leq  2p^{2}\bigg(K &\exp\Big(-n\epsilon^{2}/18K^{2} \Big) \\
 & + \exp\Big( -n\epsilon^{2}/72\Big) + 
K\exp\Big( -n\epsilon^{2}/72K^{2}\Big)\bigg).
\end{split}
\end{equation}
Now we prove $ii.$ Suppose that $\bm S\not \subseteq \hat{\bm S}$, then there exists $(j,l) \in \bm S$ such that $\hat{w}_{j,l}\leq cn^{-r} $. Therefore using condition (C2) we obtain $\big|\hat{w}_{j,l} - w_{j,l}\big|>cn^{-r}$. Hence, one has
\begin{align*}
	\Big\{\bm S \not\subseteq\hat{\bm S} \Big\} &\subseteq  \Big\{ \big|\hat{w}_{j,l} - w_{j,l}\big|>cn^{-r}, for\ a\ certain\ (j,l) \in \bm S \Big\} \\ &\subseteq \Big\{\max_{(j,l) \in \bm S}\big|\hat{w}_{j,l} - w_{j,l}\big|>cn^{-r} \Big\}.
\end{align*}
Therefore
\begin{align*}
	\mathbb{P}\Big(\bm S \not\subseteq\hat{\bm S}  \Big)
	&\leq \mathbb{P}\Big(\max_{(j,l) \in \bm S}\big|\hat{w}_{j,l} - w_{j,l}\big|>cn^{-r}\Big)  \nonumber \\
	&\leq 2s\bigg(bn^{d} \exp\Big(-c^{2}n^{1-2(r+d)}/18b^{2} \Big) \\ & \ \ \ \ \ \ \ \ \ + \exp\Big( -n^{1-2r}/72\Big) + 
	bn^{d}\exp\Big( -c^{2}n^{1-2(r+d)}/72b^{2}\Big)\bigg),
\end{align*}
where we have used (\ref{c.5}) and condition (C3) in the last line. \\
Finally, we get the sure screening property
\begin{equation*}
\begin{split}
\mathbb{P}\Big(\bm S \subseteq\hat{\bm S}  \Big) \geq 1 - 2&s\bigg(bn^{d} \exp\Big(-c^{2}n^{1-2(r+d)}/18b^{2} \Big) \\
 &\quad + \exp\Big( -n^{1-2r}/72\Big) + 
bn^{d}\exp\Big( -c^{2}n^{1-2(r+d)}/72b^{2}\Big)\bigg).
\end{split}
\end{equation*}
\begin{flushright}
	$\square$
\end{flushright}

\section{Proof of Theorem 2} \label{appD}
One has
\begin{align}\label{eq10}
&\mathbb{P}\bigg[\Big(\min\limits_{j,l\in \bm S}\hat{w}_{j,l} - \max\limits_{j,l\in \bm S^{c}}\hat{w}_{j,l}\Big) < c_3/2\bigg] \nonumber  \\
\leq& \mathbb{P}\bigg[\Big(\min\limits_{j,l\in \bm S}\hat{w}_{j,l} - \max\limits_{j,l\in \bm S^{c}}\hat{w}_{j,l}\Big) - \Big(\min\limits_{j,l\in \bm S}w_{j,l} - \max\limits_{j,l\in \bm S^{c}}w_{j,l}\Big) < -c_3/2\bigg] \nonumber  \\
\leq& \mathbb{P}\bigg[\Big|\Big(\min\limits_{j,l\in \bm S}\hat{w}_{j,l} - \max\limits_{j,l\in \bm S^{c}}\hat{w}_{j,l}\Big) - \Big(\min\limits_{j,l\in \bm S}w_{j,l} - \max\limits_{j,l\in \bm S^{c}}w_{j,l}\Big)\Big| > c_3/2\bigg] \nonumber  \\
\leq& \mathbb{P}\bigg[2\max\limits_{1\leq j,l \leq p}\Big|\hat{w}_{j,l}- w_{j,l}\Big|> c_3/2 \bigg] \nonumber  \\
\leq& 2p^{2}\bigg(K \exp\big(-nc_{3}^{2}/288K^{2} \big) + \exp\big( -nc_{3}^{2}/1152\big) + 
K\exp\big( -nc_{3}^{2}/1152K^{2}\big)\bigg)   \nonumber  \\
\leq& 2Kp^{2}\bigg(\exp\big(-nc_4\big) + 2\exp\big(-nc_4/K^{2}\big)\bigg),
\end{align}
with $c_4= c_3^{2}/1152$.\\
First, conditions ($C3$) and ($C4$) imply that $K^2\log(p^2)/n = o\big(1\big)$ which in turn implies that $p^2 \leq \exp\big(\dfrac{c_4}{2}n\big)$. Hence, one has
\begin{align}\label{eq11}
Kp^{2}\exp \Big(-c_4n\Big) &\leq \exp\bigg(\log(K) + \dfrac{c_4}{2}n - c_4n\bigg)  \nonumber \\
&\leq  \exp\bigg(\log\big(K\big) - \dfrac{c_4}{2}n \bigg) \nonumber  \\
&\leq  \exp\bigg(\log\big(K\big) - 4\log\big(n\big)\bigg) \nonumber  \\
&\leq \exp\bigg(2\log\big(n\big) - 4\log\big(n\big)\bigg) \nonumber  \\
&\leq n^{-2}.
\end{align}
Secondly, $K^2\log\big(p^2\big)/n = o\big(1\big)$ implies that $p^2 \leq \exp\big(\dfrac{c_4}{2}n/K^2\big)$, and $K^2\log(n)/n = o\big(1\big)$ implies that $4\log n \leq \dfrac{c_4}{2}n/K^2$. Hence, one has
\begin{align}\label{eq12}
Kp^{2}\exp\Big(-c_4n/K^{2}\Big) &\leq \exp \bigg( log\big(K\big) + \dfrac{c_4}{2}n/K^2 -c_4n/K^{2} \bigg)  \nonumber  \\
&\leq  \exp\bigg(\log\big(K\big) - \dfrac{c_4}{2}n/K^2 \bigg) \nonumber  \\
&\leq  \exp\bigg(\log\big(K\big) - 4\log\big(n\big)\bigg) \nonumber  \\
&\leq \exp\bigg(2\log\big(n\big) - 4\log\big(n\big)\bigg) \nonumber  \\
&\leq n^{-2}.
\end{align}
Combining (\ref{eq10}), (\ref{eq11}) and (\ref{eq12}), we get 
\begin{equation*}
\mathbb{P}\bigg\{\Big(\min\limits_{j,l\in \bm S}\hat{w}_{j,l} - \max\limits_{j,l\in \bm S^{c}}\hat{w}_{j,l}\Big) < c_3/2\bigg\} \leq  6 n^{-2}.
\end{equation*}
Hence, for some $n_0$ we have
\begin{equation*}
\sum_{n=n_0}^{\infty} \mathbb{P}\bigg\{\Big(\min\limits_{j,l\in \bm S}\hat{w}_{j,l} - \max\limits_{j,l\in \bm S^{c}}\hat{w}_{j,l}\Big) < c_3/2\bigg\} < +\infty.
\end{equation*}
Consequently, using \emph{Borel Cantelli} lemma we find the desired result
\[
\liminf\limits_{n\rightarrow\infty} \bigg\{\min\limits_{j,l\in \bm S}\hat{w}_{j,l} - \max\limits_{j,l\in \bm S^{c}}\hat{w}_{j,l} \bigg\} > 0\ \ a.s.
\]
\begin{flushright}
	$\square$
\end{flushright}
\newpage

\bibliographystyle{abbrvnat}
\bibliography{KIF-Bibliography}

\begin{figure}
	\begin{center}
		\includegraphics[scale=0.6]{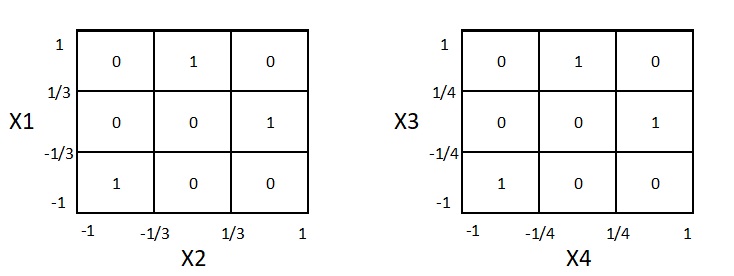}  
	\end{center}
	\caption{Relationship between the significant features and the outcome $Y$.}
	\label{figure1}
\end{figure}

\begin{figure}
	\begin{center}
		\includegraphics[scale=0.5]{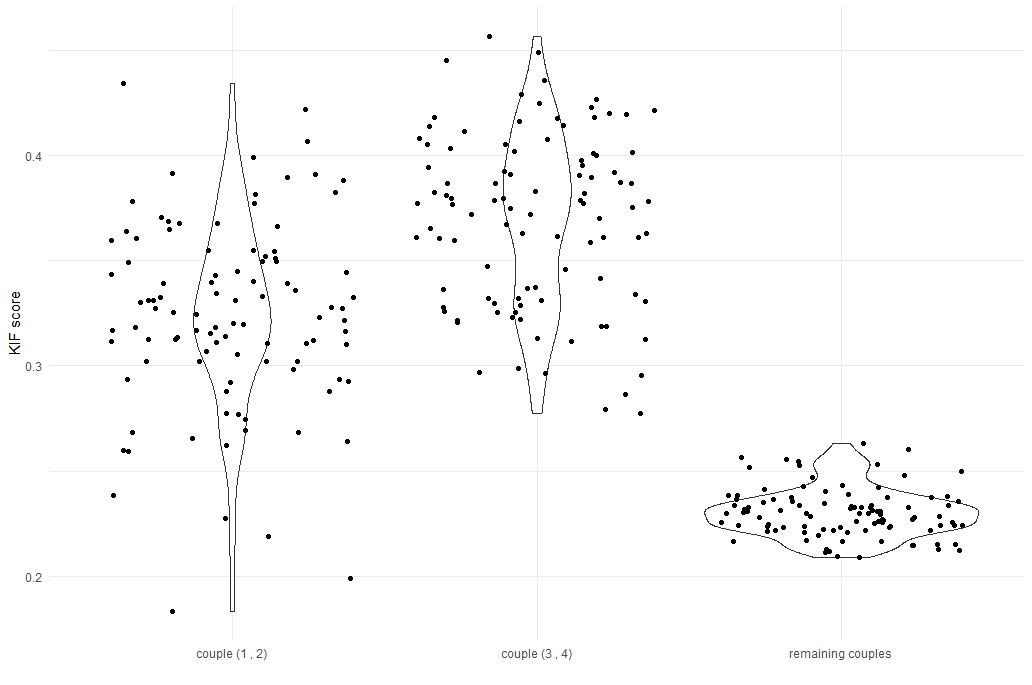}  
	\end{center}
	\caption{The KIF scores of couples $(X_1,X_2)$ and $(X_3,X_4)$ compared to the maximum KIF score of the other couples.}
	\label{figure2}
\end{figure}

\begin{table}
	\begin{center} 
		\begin{tabular}{|r| r r r r r|}
			\toprule[1pt] 
			Models & KIF & JCIS & BCOR-SIS & IP-SIS & SODA  \\ \midrule[1pt] \midrule[1pt]
			Model 1 & 1& 0.05 & 1 & 1 & 0.66 \\ 
			Model 2 & 0.89 & 0.90 & 0.99 & 0.75 & 1 \\ 
			Model 3 & 0.98 & 0.98 & 0.17 & 0.73 & 0.21 \\ 
			Model 4 & 1 & 1 & 1 & 1 & 0 \\ \bottomrule[1pt]
		\end{tabular} 
		\caption{Selection rate of couple $(X_1,X_2)$ in the four scenarios of Setting 1.}
	\end{center}
	\label{Table1}
\end{table}

\begin{table}
	\begin{center} 
		\begin{tabular}{|r |r r r r r|}
			\toprule[1pt] 
			Models & KIF & JCIS & BCOR-SIS & IP-SIS & SODA \\ \midrule[1pt] \midrule[1pt]
			Model 1 & 1 & 0.01 & 1 & 0.78 & 1 \\ 
			Model 2 & 0.89 & 0.01 & 1 & 0.56 & 0.97 \\ 
			Model 3 & 0.98 & 0.02 & 0.11 & 0.18 & 0.10 \\ 
			Model 4 & 1 & 0.09 & 1 & 0.52 & 0 \\ \bottomrule[1pt]
		\end{tabular} 
		\caption{Selection rate of couple $(X_1,X_2)$ in the four scenarios of Setting 2.}
	\end{center}
	\label{Table2}
\end{table}

\begin{table}
	\begin{center} 
		\begin{tabular}{|r r r r|}
			\toprule[1pt] 
			Couples & KIF & JCIS & SODA \\ \midrule[1pt] \midrule[1pt]
			$(X_1,X_2)$ & 0.9 & 0.71 & 0\\ 
			$(X_3,X_4)$ & 1 & 1 & 0\\ \bottomrule[1pt]
		\end{tabular} 
		\caption{Selection rate of couples $(X_1,X_2)$ and $(X_3,X_4)$ in Setting 3.}
	\end{center}
	\label{Table3}
\end{table}

\begin{table}
	\begin{center} 
		\begin{tabular}{|r r r r|}
			\toprule[1pt] 
			Couples & KIF & JCIS & SODA\\ \midrule[1pt] \midrule[1pt]
			$(X_1,X_2)$ & 0.89 & 0.71 & 0\\ 
			$(X_3,X_4)$ & 0 & 0 & 0\\ \bottomrule[1pt]
		\end{tabular} 
		\caption{Selection rate of couples $(X_1,X_2)$ and $(X_3,X_4)$ in Setting 4. $(X_1,X_2)$ is an important couple for $Y$; $(X_3,X_4)$ is a non relevant pair for $Y$.}
	\end{center}
	\label{Table4}
\end{table}

\begin{table}
	\begin{center} 
		\begin{tabular}{|c|c|c|c|c|}
			\toprule[1pt]
			\multirow{2}{*}{} & \multicolumn{4}{c|}{j} \\ \cmidrule[1pt](r){2-5}
			$\theta_{kj}$ & 1 & 2 & 3 & 4 \\ \midrule[1pt] \midrule[1pt]
			$k = 0$ & 0.3 & 0.4 & 0.5 & 0.3 \\
			$k = 1$ & 0.95 & 0.9 & 0.9 & 0.95 \\ \bottomrule[1pt]
		\end{tabular} %
		\caption{Probability parameter values used to generate the four relevant couples of features in Setting 5.}
	\end{center}
	\label{Table5}
\end{table}

\begin{table}
	\begin{center} 
		\begin{tabular}{|r|r|r r r r|}
			\toprule[1pt]
			Proportions & Couples & KIF & JCIS & PC-SIS & SODA \\ \midrule[1pt] \midrule[1pt]
			\multirow{4}{*}{$\pi_0 = 0.5$ \& $\pi_1 = 0.5$} & ($X_1, X_2$) & 1 & 0.18 & 0.61 & 0.26\\ 
			& ($X_3, X_4$) & 1 & 0.43 & 0.80 & 0.08\\ 
			& ($X_5, X_6$) & 0.98 & 0.49 & 0.78 & 0.04\\ 
			& ($X_7, X_8$) & 1 & 0.16 & 0.52 & 0.23\\ \midrule[1pt]
			\multirow{4}{*}{$\pi_0 = 0.7$ \& $\pi_1 = 0.3$} & ($X_1, X_2$) & 0.87 & 1 & 0.25 & 0.24\\ 
			& ($X_3, X_4$) & 0.87 & 1 & 0.46 & 0.13\\ 
			& ($X_5, X_6$) & 0.87 & 1 & 0.55 & 0.10\\ 
			& ($X_7, X_8$) & 0.87 & 1 & 0.20 & 0.24\\ \midrule[1pt]
			\multirow{4}{*}{$\pi_0 = 0.3$ \& $\pi_1 = 0.7$} & ($X_1, X_2$) & 0.87 & 0.22 & 0.37 & 0.20\\ 
			& ($X_3, X_4$) & 0.66 & 0.03 & 0.63 & 0.12\\ 
			& ($X_5, X_6$) & 0.62 & 0.01 & 0.70 & 0.01\\ 
			& ($X_7, X_8$) & 0.80 & 0.18 & 0.27 & 0.23\\ \bottomrule[1pt]
			
		\end{tabular} 
		\caption{Selection rate of the four relevant couples for the three scenarios of Setting 5.}
	\end{center}
	\label{Table6}
\end{table}

\begin{table}
	\begin{center} 
		\resizebox{\columnwidth}{!}{%
			\begin{tabular}{|l l l|}
				\toprule[1pt] 
				Dataset & N.groups & Description \\ \midrule[1pt] \midrule[1pt]
				Alon & 2 &  Colon Cancer \cite{Alon1999}. It consists of 2000 features and 62 observations. \\ 
				Shipp & 2 & Lymphoma \cite{Shipp2002}. It consists of 6817 features and 58 observations.\\
				West & 2 & Breast Cancer \cite{West2001}. It consists of 7129 features and 49 observations.\\ 
				Yeoh & 6 & Leukemia \cite{Yeoh2002}. It consists of 7129 features and 49 observations.\\ \bottomrule[1pt]
			\end{tabular} %
		}
		\caption{Real datasets description.}
	\end{center}
	\label{Table7}
\end{table}

\begin{table}
	\begin{center} 
		\resizebox{\columnwidth}{!}{%
			\begin{tabular}{|r|r| r r r r r r|}
				\toprule[1pt]
				data & \multicolumn{7}{c|}{Couples and their correspondent \emph{p-values}}  \\ \midrule[1pt] \midrule[1pt]
				\multirow{2}{*}{Alon} & \emph{Couple} & (265 , 1129) & (548 , 1129) & (893 , 1129) & (704 , 859) & (4 , 338) & (324 , 859)  \\ \cmidrule[1pt](r){2-8}
				& \emph{p-value} & 0.00003 & 0.00003 & 0.00001 & 0.00003 & 0.00000 & 0.00003 \\ \midrule[1pt]
				\multirow{2}{*}{Shipp} & \emph{Couple} & (34 , 1889) & (6 , 582) & (7 , 582) & (7 , 3192) & (1773 , 3601) & (3975 , 4296)\\ \cmidrule[1pt](r){2-8}
				& \emph{p-value} & 0 & 0 & 0 & 0 & 0 & 0 \\ \midrule[1pt]
				\multirow{2}{*}{West} & \emph{Couple} & (3821 , 4439) & (4331 , 5596) & (5060 , 5169) & (3276 , 3724) & (3535 , 4915) & (3535 , 5594)\\ \cmidrule[1pt](r){2-8}
				& \emph{p-value} & 0 & 0 & 0 & 0 & 0 & 0.00001 \\ \midrule[1pt]
				\multirow{2}{*}{Yeoh} & \emph{Couple} & (115 , 226) & (13 , 577) & (107 , 115) & (261 , 435) & (859 , 1091) & (13 , 136)\\ \cmidrule[1pt](r){2-8}
				& \emph{p-value} & 0 & 0 & 0 & 0 & 0 & 0 \\ \bottomrule[1pt]
			\end{tabular} %
		}
		\caption{First five selected couples and their associated p-values, using KIF, for each dataset: the couples are ranked from left to right with respect to the descending order of their KIF scores.}
	\end{center}
	\label{Table8}
\end{table}

\end{document}